\documentstyle[12pt]{article}

\def\be{\begin{equation}}
\def\en{\end{equation}}
 
\begin{document}
\begin{titlepage}
\baselineskip = 25pt
\begin{center}
 
{\Large\bf SMALL ANGULAR SCALE SIMULATIONS}
{\Large\bf OF THE MICROWAVE SKY}

\vspace{.5 cm}
{D. S\'aez $^1$, E. Holtmann $^{2}$, and G. F. Smoot $^2$}
\small
 
$^1$ Departamento de Astronom\'{\i}a y Astrof\'{\i}sica,
Universidad de Valencia, Burjassot, Valencia, Spain\\
 
$^{2}$ Lawrence Berkeley Laboratory, Space Science Laboratory,
CfPA, Bldg 50-204,
University of California, Berkeley CA 94720.
 
\footnotesize
e-mail: DIEGO.SAEZ@UV.ES\\
 
\end{center}
 
%\vspace {2 cm}
\normalsize
\begin{abstract}
 
We describe and compare
two types of microwave sky simulations which are
good for small angular scales.
The first type uses expansions in
spherical harmonics, and the second one is based on plane
waves and the  Fast Fourier
Transform. The angular power spectrum is
extracted from maps corresponding to both types of simulations,
and the resulting
spectra are appropriately
compared. In this way,
the features and usefulness of Fourier simulations
are pointed out.   For $\ell \geq 100$,
all the simulations lead to similar accuracies; however,
the CPU cost of Fourier simulations is $\sim 10$ times smaller than
that for spherical harmonic simulations.
For $\ell \leq 100$, the simulations based on
spherical harmonics seem to be preferable.
\end{abstract}
 
{\em Subject headings:} cosmic microwave background---cosmology:
theory---large-scale structure of the universe

\end{titlepage}
 
\section{Introduction}

There are important questions which could be addressed through
simulated maps of the microwave sky. How well can we recover
the angular power spectrum  of the cosmic microwave background (CMB)
and the cosmological parameters from maps of a given experiment,
which can have holes and regions with non-uniform sampling?
How well can we subtract foregrounds from a multifrequency map?
An important issue for small angular scale simulations is
their great computational cost.
It is important to have fast and accurate methods to do simulations.
This paper is concerned with the comparison of two of these
methods.
 
Large-scale simulations of the microwave sky
are based on the expansion of the temperature contrast in spherical harmonics,
\be
\frac {\delta T}{T}(\theta,\phi)=\sum_{l=1}^{l_{max}}\sum_{m=-l}^{m=+l}
a_{lm}Y_{lm}(\theta,\phi).
\en
Typically $\l_{max} \leq 40$ in COBE-like maps, since
the number of coefficients, $a_{lm}$, to be calculated is $(l_{max} + 1)^2$,
and there are questions of numerical accuracy in high order calculations.
In order to make high resolution maps
--- resolutions $\sim 10^{\prime}$---
$l_{max}$ must be of order $10^{3}$
so that the number of coefficients to be calculated is of order $10^{6}$;
moreover, the $10^{6}$ spherical harmonics must be evaluated accurately
in roughly $10^{6}$ locations in a full-sky map ($10^{4}$ locations in
a $20^{\circ} \times 20^{\circ} $ field).
Thus, for high angular resolution,
the spherical harmonic expansion becomes
difficult even with modern computing power. Although simulations
based on spherical harmonics are feasible (Hinshaw, Bennett \& Kogut 1995;
Kogut, Hinshaw \& Bennett 1995; Jungman et al. 1995),
the Fast Fourier Transform (FFT) is shown to be useful
in order to make simulations efficiently on small angular scales.
Hereafter, simulations using spherical harmonics are called
``SH simulations'', while those based on plane waves
are called ``FFT simulations''.
The main goal of this paper is the comparison of the accuracies
and the CPU costs corresponding to SH and FFT simulations.
Both approaches use the
$C_{\ell} \equiv \sum_{m = -\ell}^{m = \ell} |a_{\ell m}|^2   /(2\ell + 1) $
quantities corresponding to the model described in Sec.\ 2.
The method used to extract the angular power spectra
from the simulated maps is described
in Sec.\ 3.
FFT simulations are described in Sec.\ 4, while SH simulations
are presented in Sec.\ 5.
Comparisons between FFT and SH simulations are
given in Sec.\ 6 and, finally, Sec.\ 7 contains the main results and a
general discussion.
 
\section{The model}
 
All the simulations presented in this paper
are based on a single cold dark matter (CDM)
model for large scale structure formation.
In this simple CDM model,
no reionization modifies the CMB anisotropies,
the background is flat ($\Omega_{0}=1$),
the cosmological constant vanishes,
scalar fluctuations are Gaussian, their spectrum
is scale-invariant, and tensor fluctuations are absent.
This model is hereafter called the minimal CDM model.
The free parameters of this model are:
the amplitude of the angular power spectrum of the CMB,
the density parameter of the baryonic matter $\Omega_{_{B}}$
and the reduced Hubble constant $h$.
The parameters $\Omega_{_{B}}$
and $h$ are involved in the transfer function; their values
are assumed to be
$0.03$ and $1/2$, respectively.
 
In the above model,
the normalization of the angular power spectrum is fixed by
to rms quadrupole of the CMB anisotropy generated by scalar modes
by using the estimator $Q_{rms-PS}$ found by fitting the
observed temperature fluctuations assuming
a scale-invariant primordial density power spectrum.
In the absence of tensor modes,
experiments measuring at large angular scales, such as COBE (Smoot
et al. 1992, Bennett et al. 1992, Wright et al. 1992)
and TENERIFE (Hancock et al. 1994), provide estimations of $Q_{rms-PS}$.
The $C_{\ell}$ coefficients
have been taken from Sugiyama (1995) and
renormalized according to the four-year
COBE data ($Q_{rms-PS} \simeq 18 \ \mu K$, Gorski et al. 1996).
All the simulations performed in this paper are based on
these $C_{\ell}$ coefficients.

\section{Obtaining the angular power spectrum from simulated maps}
 
In this section, we describe the method used in order to extract the
angular power spectrum from a given map.
Let us give some useful definitions.
 
The autocorrelation function can be defined as
$C(\theta)=C_{\sigma=0}(\theta)$, where
\be
C_{\sigma}(\theta)=\left \langle
\left( \frac {\delta T}{T}\right)_{\sigma} ({\vec {n}}_{1})
\left( \frac {\delta T}{T}\right)_{\sigma} ({\vec {n}}_{2})
\right \rangle \ .
\en
The angle between the unit vectors $\vec {n}_{1}$ and $\vec {n}_{2}$
is $\theta$. The
angular brackets stand for the mean over many full realizations
of the CMB sky. The quantity $(\frac {\delta T}{T})_{\sigma} ( \vec {n})$
is the temperature contrast in the direction $ \vec {n} $ after
smoothing with a Gaussian beam described by
$\sigma=0.425 \theta_{_{FWHM}}$.
 
Function $C_{\sigma}(\theta)$ can be expanded in the following
form:
\be
C_{\sigma}(\theta)= \frac {1}{4 \pi} \sum_{\ell =2}^{\infty}
(2\ell +1)C_{\ell} P_{\ell}(\cos {\theta}) e^{(- \ell + 0.5)^{2}
\sigma^{2}}.
\en
where the $C_{\ell}$ quantities define the power spectrum of the
CMB. These quantities have been calculated in many
theoretical models of structure formation.
 
From Eq. (3) one easily obtains the relation:
\be
C_{\ell}(\sigma ) = e^{(- \ell + 0.5)^{2}
\sigma^{2}} C_{\ell}
=\frac {32 \pi^{3}} {(2 \ell +1)^{2}}
\int_{0}^{\pi} C_{\sigma}(\theta) P_{\ell} (\cos {\theta})
\sin {\theta} d \theta \ .
\en
 
Given a simulated map, the function
$C_{\sigma}(\theta)$ is obtained by using Eq. (2).
For each value of $\theta$, many pairs
of direction vectors ($\vec {n}_{1}$, $\vec {n}_{2}$)---forming a fixed
angle $\theta$---are randomly chosen.
To be more precise, the direction vector $\vec {n}_{1}$ is chosen to point
at a random node of the map---where the temperature
is known---while the direction $\vec {n}_{2}$ is randomly placed
on a cone centered on the node corresponding to
$\vec {n}_{1}$. The half-width of this cone is $\theta$.
The temperature corresponding to the direction
$\vec {n}_{2}$ is obtained by interpolating between neighboring
nodes. As discussed below,
mathematical interpolation introduces
only a small error in the resulting spectrum for
values of $\theta$ lying in the interval
($\theta_{min}$, $\theta_{max}$), where $\theta_{min}$
is the angle separating two neighboring nodes and
$\theta_{max}$ is an angle to be experimentally obtained.
This angle must be: (a) smaller than the size of the
simulated region in  order to allow us to place
a great number of pairs
($\vec {n}_{1}$, $\vec {n}_{2}$) inside it, and (b) large enough
to include as many scales as possible.

Since we know
the values of the function $C_{\sigma}(\theta)$ only interval
($\theta_{min}$, $\theta_{max}$), we cannot compute
$C_{\ell}(\sigma)$.  Instead, we
have calculated the quantities
\be
D_{\ell}(\sigma )
=\frac {32 \pi^{3}} {(2 \ell +1)^{2}}
\int_{\theta_{min}}^{\theta_{max}} C_{\sigma}(\theta)
P_{\ell} (\cos {\theta}) \sin {\theta} d \theta \ .
\en
These quantities are to be compared with the
corresponding theoretical ones, which can be obtained as
follows: First, from Sugiyama's
$C_{\ell}$ coefficients of the minimal CDM
model and from Eq.\ (3) (with the sum extended
from $\ell=2$ to $\ell=1100$), the autocorrelation function is estimated
in the interval ($\theta_{min}$, $\theta_{max}$).
This theoretical autocorrelation function plus Eq.\ (5) then
allows us to compute the theoretical values of
$D_{\ell}(\sigma )$.
 
It is hereafter said that the quantities
$D_{\ell}(\sigma )$ define the {\em modified
power spectrum}, which is different from the true
angular power spectrum given by the
coefficients $C_{\ell}(\sigma )$.
 
In all the simulations we are considering in this paper, the
following values have been selected: $\theta_{min}=5^{\prime}$,
$\theta_{max}=4.5^{\circ}$ and $\theta_{_{FWHM}}=10^{\prime}$.
These values are similar to those of modern projects for
the detection of small angular scale anisotropies.
 
A very accurate estimation of the mean involved in Eq.\ (2) requires
multiple realizations of
the microwave sky.
In the case of a single full realization,
a certain error in this mean---producing an error in
$C_{\sigma}(\theta)$---is unavoidable. This error is usually
measured by the cosmic sample variance. The smaller the sky coverage,
the greater the errors in the resulting autocorrelation functions
and spectra (Scott, Srednicki \& White, 1994).
As shown below, the spectra resulting from a small
$20^{\circ} \times 20^{\circ}$ coverage
are rather different than the theoretical ones.
Although coverages greater than $20^{\circ} \times 20^{\circ}$
cannot be achieved by using FFT simulations, we can use multiple
disconnected $20^{\circ} \times 20^{\circ}$ regions
in order to improve on the estimate of
the mean involved in Eq.\ (2), for values of
$\theta$ smaller than $\theta_{max}$.
In other words, FFT simulations
become useful as a result of the fact that
various independent small coverages of the sky can be
used in order to obtain a good estimate of the
autocorrelation function for small angular scales; however,
neither a
$20^{\circ} \times 20^{\circ}$ map nor an ensemble of them
has good information about angular correlations
larger than $\theta_{max}$.

\section{Simulations based on the FFT}
 
Our FFT simulations are based on the use of the
standard $C_{\ell}$ coefficients, which have already been
calculated for many cases.
As pointed out by Bond and Efstathiou (1987), a FFT simulation
is only possible in
the case of anisotropies corresponding to large $\ell$-values and
small regions of the sky. A quantitative study of the validity
and usefulness of this type of simulation has not previously
been presented. In order to
do this study, we compare results from FFT simulations with
results from SH simulations.
 
The temperature contrast in the simulated region can be obtained
from the following formula:
\be
\frac {\delta T}{T}(\theta, \phi)=
\sum_{s_{1},s_{2}=-N}^{N}  D(\ell_{1},\ell_{2})
e^{-i(\theta \ell_{1}+ \phi \ell_{2})}
\en
where $\ell_{1}=2 \pi s_{1}/ \Lambda$ and
$\ell_{2}=2 \pi s_{2}/ \Lambda$, $\Lambda$ being the angular size
of the elemental square. The distribution
of the quantities
$D(\ell_{1},\ell_{2})$ is assumed to be Gaussian,
with zero mean and with variance
$C_{\ell}/ \Lambda^{2}$,
where $\ell=(\ell_{1}^{2}+\ell_{2}^{2})^{1/2}$.
The quantities $D(\ell_{1},\ell_{2})$ satisfy the relation
$D(- \ell_{1}, - \ell_{2}) = D^{*}(\ell_{1},\ell_{2})$.
 
The use of plane waves requires
negligible spatial curvature in the simulated regions. This
restricts our simulations to small parts
of the microwave sky.
In this paper,
it is verified that a region of $20^{\circ} \times 20^{\circ}$
($\sim$ 1\% of the total sky)
can be successfully mapped by using the FFT.
The value $\theta_{max} = 4.5^{\circ}$ is marginally compatible with
the size of this region (many pairs
($\vec {n}_{1}$, $\vec {n}_{2}$) forming an angle of
$4.5^{\circ}$ can be located inside
it). An angular scale of $4.5^{\circ}$
corresponds to $\ell = 40$ and, consequently,
the spectra obtained from $20^{\circ} \times 20^{\circ}$ FFT simulations is
only estimated for $\ell > 40$. In practice, the accuracy in the estimation
of the power spectrum from these simulations seems to be good for
$\ell > 100$ (see Sec.\ 6).
 
In a $20^{\circ} \times 20^{\circ}$  region, the assumption of
flatness produces a deformation of the true maps, but
this deformation is not expected to hide their main features.
We remind the reader that only correlations
at scales smaller than $4.5^{\circ}$ are being estimated and,
on these scales, the curvature is clearly negligible.
 
Since the FT involves periodic boundary conditions,
our simulations are not physically significant at all points
of the
elemental square. Only the simulation of the central region is
physically admissible.  On account of this fact, Fourier
transforms are
performed in a large $\sim 40^{\circ} \times 40^{\circ}$
square ($\Lambda = 40^{\circ}$)
and the resulting simulation is truncated to the
central $\sim 20^{\circ} \times 20^{\circ}$ region.
The curvature could be important if we wanted to get a
physically significant mapping of the
large square; however,
the large square is only an auxiliary element without
any physical significance.
The number of points on each edge of the large square is
taken to be 2N=512.
This choice corresponds to $\theta_{min} \sim 5^{\prime}$.
Simulations using
an $\sim 80^{\circ} \times 80^{\circ}$ square with 1024
points per edge yield similar results;
hence, the use of more than 512 points per edge
is not necessary.
 
On an IBM 30-9021 VF, the CPU cost of our FORTRAN code
is $\sim 5.5$ minutes per simulation.
 
Figure  (1) shows a FFT simulation of a
$\sim 20^{\circ} \times 20^{\circ}$ region.
No unusual features are apparent near the edges.
The numbers appearing in this Figure are the minimum and maximum values
of the temperature contrast in our simulation.
Only a few points of the simulation approach these values.

\section{Simulations based on spherical harmonics}
 
Our
SH simulations are based on Eq.\ (1)
with $\ell_{max} = 1100$ and $\ell_{min} = 40$.
The  $a_{\ell m}$ coefficients have been generated as statistically
independent random numbers with
variance $\langle  |a_{\ell m}|^{2} \rangle = C_{\ell}
e^{(- \ell + 0.5)^{2} \sigma^{2}}$
and zero mean.
The spherical harmonics have been carefully calculated.
These simulations include scales smaller than those
considered in previous simulations
(see Hinshaw, Bennett \& Kogut 1995;
Kogut, Hinshaw \& Bennett 1995; Jungman et al. 1995).
The small values of $\theta_{min}$
and $\sigma$ considered in our simulations
require the use of large values of $\ell$ giving information about
small angular scales.
 
In this paper, SH simulations are performed only for comparison
with FFT simulations; for this reason,
some features of our numerical SH simulations are identical
to those of the FFT simulations of Sec.\ 4. In particular,
our SH simulations
are extended to
$20^{\circ} \times 20^{\circ}$
regions of the sky.
These regions are assumed to be uniformly
covered and the number of points per edge is 256
($\theta_{min} \sim 5^{\prime}$).
A smoothing corresponding to
$\theta_{_{FWHM}} = 10^{\prime}$ has been performed, and
the angle $\theta_{max} = 4.5^{\circ}$ has been assumed.
 
Simulations of the two kinds defined up to now have been obtained
for the minimal CDM model. The $C_{\ell}$
coefficients have been obtained and normalized as detailed in Sec.\ 2.
 
On an IBM 30-9021 VF, the CPU cost of our FORTRAN code
is $\sim 53$ minutes per simulation.
This cost is about ten times greater than that of the FFT
simulations described in Sec.\ 4.
 
SH simulations of $\sim 20^{\circ} \times 20^{\circ}$ regions look like the
FFT simulation of Fig.\ 1. Differences between SH and FFT simulations
can be noticed only by comparing the spectra obtained from them.
These comparisons are discussed in the next section.
 
\section{Results}
 
Let us now compare the theoretical $D_{\ell}(\sigma )$ quantities
with those
obtained from FFT and SH simulations.
 
In all the panels of Fig.\ (2), the solid lines
correspond to
the modified theoretical spectrum and the dashed lines
show the modified spectrum obtained from simulations.
The left (right) panels
of this figure show the modified spectra
extracted from
FFT (SH)  simulations.
The top, intermediate, and bottom panels exhibit the modified
spectra obtained from 1, 13, and 52 small simulations of
$20^{\circ} \times 20^{\circ}$ regions, respectively.
The total area covered by the 1, 13, and 52 small simulations
is 400, 5400, and 21600 ( $\sim  \frac {1} {2}$ of
the full sky area) squared degrees, respectively.
Hereafter, $N_{s}$ stands for
the number of $20^{\circ} \times 20^{\circ}$ simulations.

In order to measure the deviations between the theoretical
and simulated spectra in each case,
the following quantities have been calculated
and presented in Table 1 : The mean, $M1$, of the quantities
$0.69 \ell (\ell +1)D_{\ell}(\sigma)
\times 10^{10}$ (column 3); the mean, $M2$, of the differences
between the theoretical and simulated values of these quantities (column 4);
the typical deviation, $\Sigma$, of the same (column 5); and
the mean, $MA$, of the absolute value of these differences
(column 6). These
quantities are estimated in appropriate $\ell$ intervals
(column 7).
 
As can be seen from the top panel of Fig.\ 2 and from
Entries 1 and 2
of Table 1 (see $MA$ and $\Sigma$),
a single $20^{\circ} \times 20^{\circ}$ map
leads to a poor estimation of the angular spectrum in
both FFT and SH simulations.
A greater sky coverage is necessary.
The intermediate panels of Fig.\ 2 and Entries 3 and 4 of Table 1
prove that the use of thirteen $20^{\circ} \times 20^{\circ}$ maps
strongly improves on the resulting spectra (see $MA$ and $\Sigma$).
The quantities $MA$ and $\Sigma$
decrease as $N_{s}$
increases.
The total decrease undergone by
MA and $\Sigma$ as $N_{s}$ goes from  1 to 13
(see the top and intermediate panels of Fig. 2 and
Entries 1--4 of Table 1)
is larger
than that appearing as $N_{s}$ increases from 13 to 52
(see the intermediate and bottom panels of Fig. 2 and
Entries 3--6 of Table 1).
The quantity $M2$ is of order $10^{-2}$ whatever the value of
$N_{s}$ may be. The absolute
value of $M2$ is always much smaller than that of $MA$; this
means that the deviations between the theoretical and
the simulated spectra are essentially oscillatory. This
fact strongly supports the idea that moderate coverages
could give good spectra after removing oscillations.
The above results can be independently obtained from both
SH and FFT simulations.
 
For $N_{s}=52$ and the $\ell$-interval ($40,100$), the
quantities $| M2 |$, $MA$, and $\Sigma$ corresponding to FFT simulations
are greater than those of SH simulations
(compare Entries 7 and 8 of Table 1). This means that,
for $\ell<100$, SH simulations are preferable
(in spite of their large CPU cost).
 
The above results have been verified in other
realizations of FFT and SH simulations corresponding to
the $N_{s}$ values 1, 13 and 52; however,
only one set of realizations
is presented (Fig.\ 2) for the sake of brevity.
 
Although this paper is devoted primarily to the comparison of FFT
and SH simulations in the absence of noise,
the effect of uncorrelated noise at the level of $27.3 \ \mu K$ in a
$20^{\circ} \times 20^{\circ}$ region has been estimated.
In order to do this, a $20^{\circ} \times 20^{\circ}$ simulation
involving only uncorrelated noise has been made.  The
spectrum has been obtained from the resulting map
following the prescriptions of Sec.\ 3.
The quantities $0.69 \ell (\ell +1)D_{\ell}(\sigma)
\times 10^{10}$ appear to be smaller than $10^{-2}$ for all $\ell$.
These values are negligible compared to those
corresponding to the fluctuations of the microwave background
(see Fig.\ (2); $0.69 \ell (\ell +1)D_{\ell}(\sigma)
\times 10^{10} \sim 1$). This proves that the effect of uncorrelated noise
in our $20^{\circ} \times 20^{\circ}$ maps can be neglected.
This fact has been verified by superimposing the above white noise map
upon a SH map. The spectra obtained from the resulting map
is practically the same as in the absence of noise.

\section{Conclusions and discussion}
 
Small scale simulations of the microwave background are
expected to be very useful in order to analyze observational
data from both current and future experiments.
For scales smaller (greater) than $\sim 1.8^{\circ}$ ($\ell \sim 100$),
the accuracy of the FFT simulations appears to be comparable to
(worse than)
that of the SH simulations. The CPU cost
of the SH simulations is about ten times greater than that of the
FFT simulations. These conclusions strongly support the
use of FFT simulations as a fast and accurate tool leading to
very good spectra in the case of small angular scales ($\ell>100$).
 
There are various sources of
the deviations between the theoretical spectrum and
the spectrum extracted from FFT simulations.
Significant contributions to these deviations are expected from
partial coverage, boundary conditions in
FFT maps, and the use of
mathematical interpolation in order to assign a
temperature contrast to the direction $\vec {n}_{2}$
(see Sec.\ 3). Such
interpolation introduces non-physical information
even for the
most sophisticated interpolation methods. An error
in the resulting spectrum associated
with this fact seems to be unavoidable. Fortunately, this
error appears to be small from $\theta_{min}$ to $\theta_{max}$.
Boundary effects associated with the FFT have been minimized by
simulating the core of an auxiliary $40^{\circ} \times 40^{\circ}$ region.
 
Uncorrelated noise does not appear to be relevant for the
coverages considered in this paper.
 
Since we are only considering
angular scales smaller that $4.5^{\circ}$, curvature effects
are expected to be negligible.
 
The main conclusions of this paper are independent of
admissible renormalizations of the $C_{\ell}$ coefficients.
 
{\large \bf Acknowledgments}
 
Numerical computations were carried out at both
the Lawrence Berkeley Laboratory and the Computational Center
of Valencia University.
D.S. wishes to thank P. Andreu for useful discussions and the
Conselleria de Cultura Educacio i Ciencia
de la Generalitat Valenciana for a grant. This work
was partially supported by the project GV-2207/94.
 
\newpage
 
%\vspace{1cm}
{\Large\bf References}
\\
Bennett, C.L. et al., 1992, ApJ, 396, L7\\
Bond, J.R. \& Efstathiou, G., 1987, MNRAS, 226, 655\\
G\'orski et al, 1996, astro-ph/9601063, submitted to ApJ. Letters\\
Hancock, S., et al., 1994, Nature, 367, 333\\
Hinshaw, G., Bennett, C.L., \& Kogut, A., 1995, ApJ, 441, L1\\
Jungman, G., Kamionkowski, M., Kosowsky, A. \& Spergel, D.N.,
1995, Astro-ph/9512139\\
Kogut, A., Hinshaw, G. \& Bennett, C.L., 1995, ApJ, 441, L5\\
Scott, D., Srednicki, M. \& White, M., 1994, ApJ, 421, L5\\
Smoot, G.F. et al., 1992, ApJ, 396, L1\\
Sugiyama, N., 1995, ApJ Supplement, 100, 281\\
Wright, E.L. et al., 1992, ApJ, 396, L13\\
 
\newpage

\begin{table}[b]
\begin{center}
TABLE 1\\
COMPARING THEORETICAL AND SIMULATED SPECTRA\\
 \begin{tabular}{ccccccc}\\
\hline
\hline
Type of &   &   &  &  &  & \\
simulations & $N_{s}$ & $M1$ & $M2$ & $\Sigma$ &
$MA$ & $\ell$-interval\\
\hline
FFT  & 1 & $1.07$ &  $-1.01 \times 10^{-2}$ &
      $7.13 \times 10^{-3}$ & $1.58 \times 10^{-1}$  & $40 - 1000$\\
SH & 1 & $1.07$ &  $-1.33 \times 10^{-2}$ &
      $6.96 \times 10^{-3}$ & $1.76 \times 10^{-1}$ & $40 - 1000$\\
FFT & 13 & $1.07$ &  $-1.13 \times 10^{-2}$ &
      $2.58 \times 10^{-3}$ & $5.90 \times 10^{-2}$ & $40 - 1000$\\
SH & 13 & $1.07$ &  $-1.85 \times 10^{-2}$ &
      $3.08 \times 10^{-3}$ & $8.00 \times 10^{-2}$ & $40 - 1000$\\
FFT & 52 & $1.07$ &  $-2.04 \times 10^{-2}$ &
      $2.25 \times 10^{-3}$ & $5.21 \times 10^{-2}$ & $40 - 1000$\\
SH & 52 & $1.07$ &  $-2.56 \times 10^{-2}$ &
      $2.30 \times 10^{-3}$ & $6.06 \times 10^{-2}$ & $40 - 1000$\\
FFT & 52 & $1.24$ &  $-5.29 \times 10^{-2}$ &
      $2.46 \times 10^{-2}$ & $1.61 \times 10^{-1}$ & $40 - 100$\\
SH & 52 & $1.24$ &  $-1.54 \times 10^{-2}$ &
      $1.05 \times 10^{-2}$ & $6.54 \times 10^{-2}$ & $40 - 100$\\
\hline
\multicolumn{7}{c}{}\\
\end{tabular}
\end{center}
\end{table}
 
\newpage
 
\begin{center}
{\bf Figure Captions}
\end{center}
 
\noindent
{\bf Fig. 1.} FFT simulation of a $20^{\circ} \times 20^{\circ}$
region of the microwave sky. Numbers correspond to the minimum and maximum
values of the temperature contrast.
\vskip 0.5cm
\noindent
{\bf Fig. 2.} Each panel shows the quantity
$0.69 \ell (\ell + 1) D_{\ell}(\sigma) \times 10^{10}$ as a function of
log($\ell$) in various cases. In all the panels, the solid
line shows the theoretical
modified power spectrum and the dashed line shows
the modified spectrum extracted from simulations.
All the modified spectra have been obtained for
$\sigma = 10^{\prime}$, $\theta_{min}=5^{\prime}$, and
$\theta_{max}= 4.5^{\circ}$.
Top, intermediate, and bottom panels correspond to the $N_{s}$ values
1, 13, and 52, respectively. Left (right) panels show
modified spectra obtained from FFT (SH) simulations.
\vskip 0.5cm
\noindent

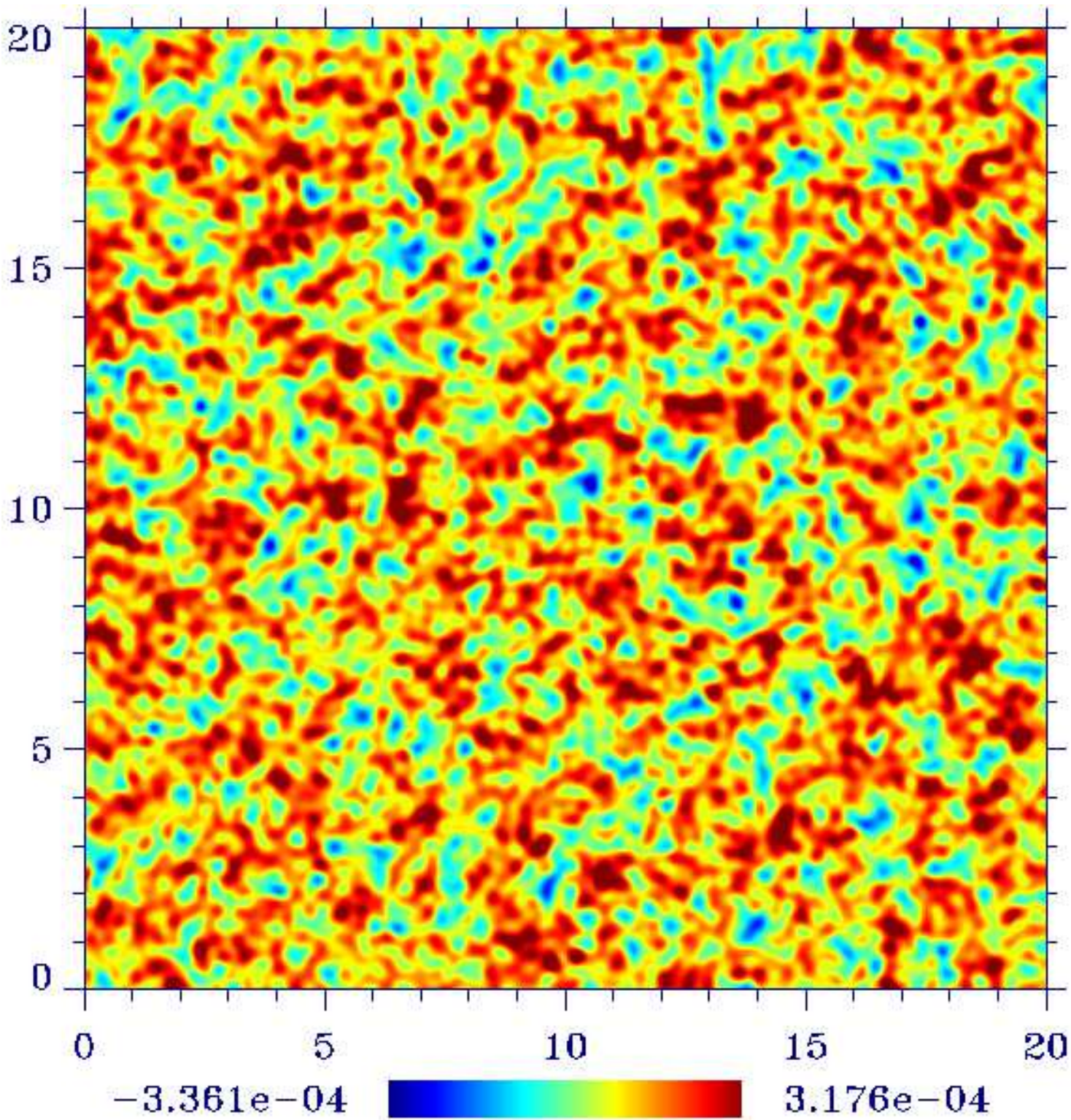
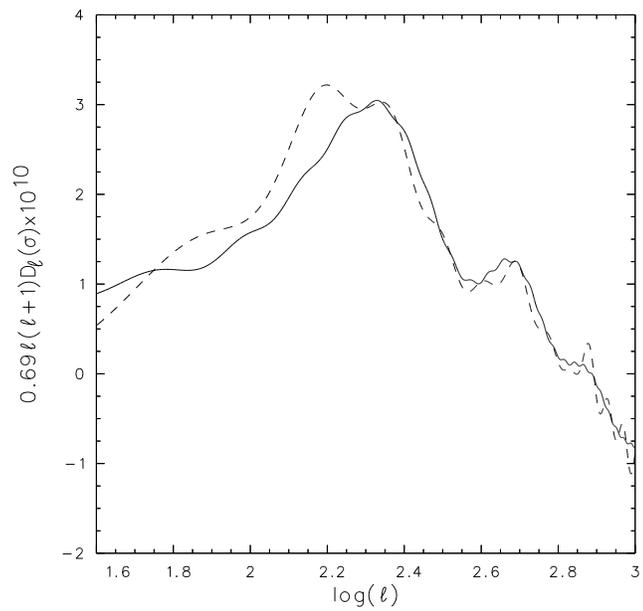
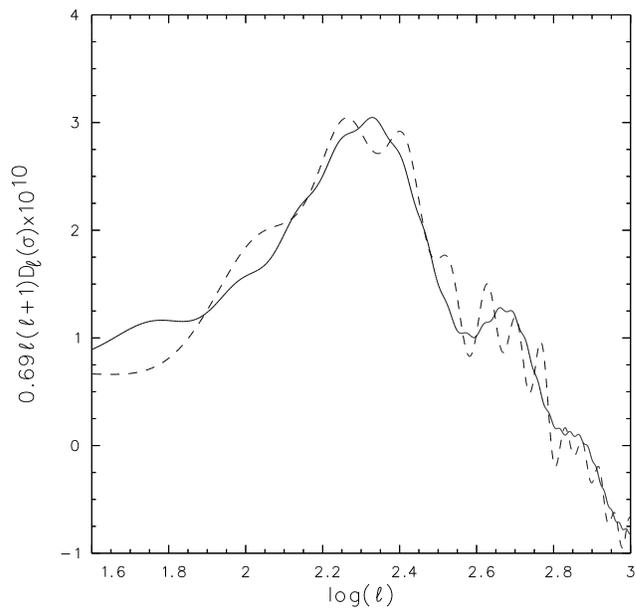
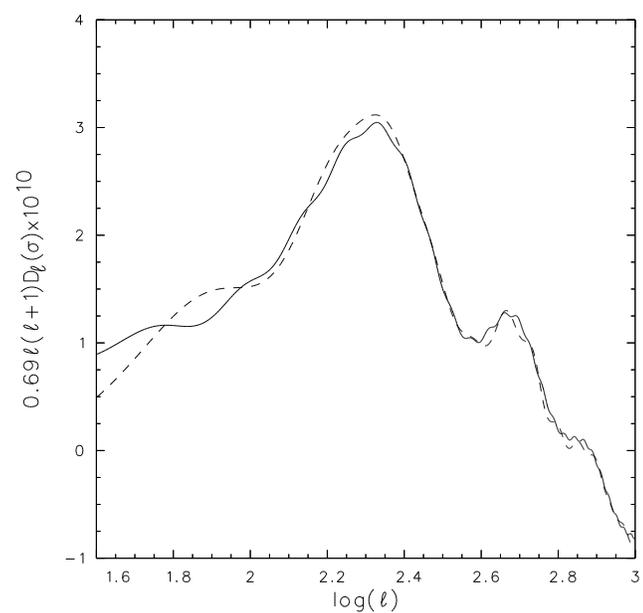
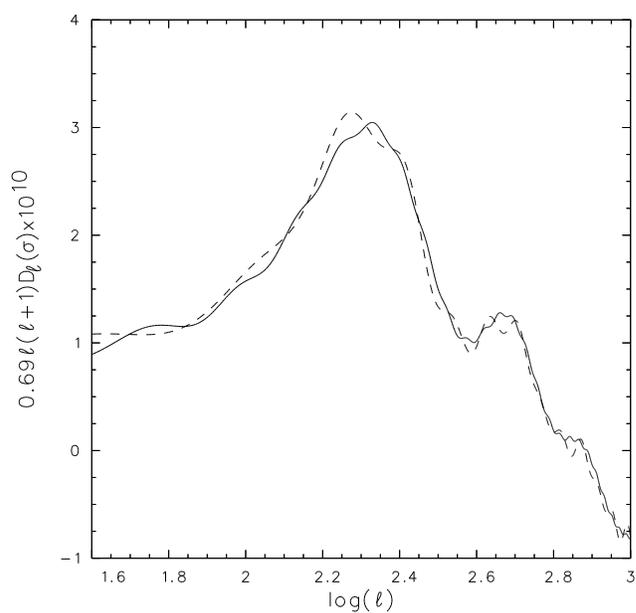
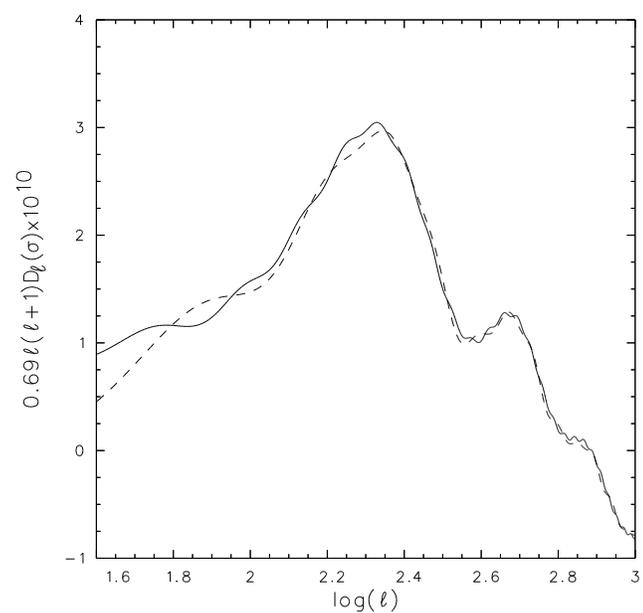
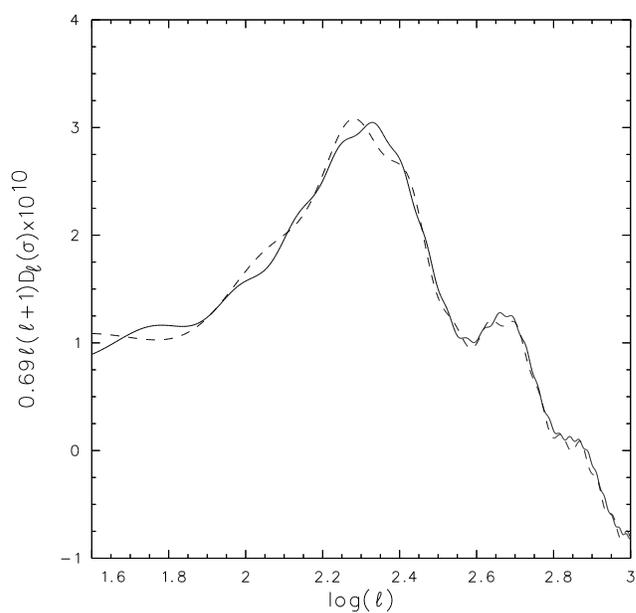
\end{document}